\documentclass[aps,prd,twocolumn,superscriptaddress,nofootinbib]{revtex4-1}

\usepackage{latexsym}
\usepackage{amsmath}
\usepackage{amssymb}
\usepackage{amsfonts}
\usepackage{bm}
\usepackage{physics}
\usepackage{slashed}

\usepackage{color}
\definecolor{purple}{rgb}{0.5,0,0.5}
\definecolor{blue}{rgb}{0.0,0,0.9}
\definecolor{prdblue}{rgb}{0.133,0.118,0.498}
\usepackage[colorlinks=true, pdfstartview=FitV, linkcolor=prdblue, citecolor= prdblue, urlcolor=prdblue]{hyperref}

\usepackage{supertabular} 
\usepackage{placeins}
\usepackage{epsfig}
\usepackage{graphicx}
\usepackage{slashed}
\begin{document}

\title{Landau–Khalatnikov–Fradkin Transformations in Reduced Quantum Electrodynamics: Perturbative and Nonperturbative Dynamics of the Fermion Propagator}

\author{Anam Ashraf}
\email[]{anam.ashraf@au.edu.pk}
\affiliation{Department of Physics, Quaid-i-Azam University, Islamabad 45320, Pakistan.}
\affiliation{Department of Physics, Air University, Islamabad, Pakistan.}

\author{Faisal Akram}
\email[]
{faisal.chep@pu.edu.pk}
\affiliation{Centre For High Energy Physics, University of the Punjab, Lahore 54590, Pakistan}

\author{M. Jamil Aslam}
\email[]{jamil@qau.edu.pk}
\affiliation{Department of Physics, Quaid-i-Azam University, Islamabad 45320, Pakistan.}

\author{Dania Rodríguez-Tzintzun}
\email[]{1717697x@umich.mx}
\affiliation{Instituto de F\'isica y Matem\'aticas, Universidad Michoacana de San Nicol\'as de Hidalgo, \\
Morelia, Michoac\'an 58040, M\'exico}

\author{Adnan Bashir}
\email[]{adnan.bashir@dci.uhu.es}
\affiliation{Departamento de Ciencias Integradas, Centro de Estudios Avanzados en Fis., Mat. y Comp., \\
Facultad de Ciencias Experimentales, Universidad de Huelva, Huelva, 21071, Spain}
\affiliation{Facultad de Ingenier\'ia, Universidad Aut\'onoma de Quer\'etaro, Quer\'etaro, Quer\'etaro 76010, M\'exico}

\author{Luis Albino}
\email[]{luis.albino.fernandez@gmail.com}
\affiliation{Departamento de Física, Universidad de Sonora,
Boulevard Luis Encinas J. y Rosales, 83000, Hermosillo, Sonora, M\'exico.}

\date{\today}

\begin{abstract}

We present a comprehensive analysis of the Landau–Khalatnikov–Fradkin transformations for the charged fermion propagator in reduced quantum electrodynamics (RQED). Starting from the propagator in a reference gauge, we perform a gauge transformation to obtain its analytical expression valid to all orders in an arbitrary covariant gauge and also applicable in a nonperturbative context.
This work complements and extends previous studies of quantum electrodynamics in various spacetime dimensions, for both massless and massive fermions. At the perturbative level, we expand the resulting expressions up to two-loop order for both massless and massive cases, and compare our results with those available in the literature wherever possible. We argue that the most suitable choice of the reference covariant gauge in RQED is $\xi=1/3$, as in this case the leading logarithmic contribution to the massless wave-function renormalization vanishes at one-loop order. This choice provides a direct connection between perturbation theory and the constraints imposed by multiplicative renormalizability on the massless fermion propagator. We also investigate the implications of the Landau–Khalatnikov–Fradkin transformations for the dynamically generated mass function of the fermion propagator. Finally, through  numerical computation, we demonstrate that both the chiral fermion condensate and the fermion pole mass are gauge-invariant quantities.

\end{abstract}

\maketitle
\section{Introduction}
Quantum electrodynamics (QED) describes a kaleidoscopic range of physical phenomena involving the interaction of charged particles — such as electrons and muons — through the exchange of photons. Among other observables, its remarkably precise predictions for the anomalous magnetic moment of the muon (see, for example, the most recent reviews~\cite{Aoyama:2020ynm,Colangelo:2022jxc,Aliberti:2025beg}) and the Lamb shift in hydrogen energy levels ~\cite{Colangelo:2022jxc,Aliberti:2025beg} make QED the most accurate and stringently tested theory in all of physics.

Beyond its central role in precision tests of the Standard Model (SM), electromagnetic interactions are essential for elucidating the internal structure of hadrons, since quarks carry electric charge and can therefore be probed electromagnetically~\cite{Aznauryan:2012ba,Aguilar:2019teb,Miramontes:2021exi,Burkert:2025coj}.
The Lagrangian for the QED has the form
\begin{eqnarray}
\mathcal{L} & = & -\frac{1}{4}F^{\mu\nu}F_{\mu\nu} -\frac{1}{2\xi}\left(\partial_\mu A^\mu\right)^2-\bar{\psi}\left(i\slashed{\partial}-m\right)\psi\notag\\
&&-e\bar{\psi}\gamma^\mu\psi A_\mu\,, \label{01}
\end{eqnarray}
where $A_\mu$, $\psi$ and $F_{\mu\nu} \equiv \partial_\mu A_\nu -\partial_\nu A_\mu$ are the gauge field, the fermion field and the electromagnetic field tensor, respectively. The strength of the coupling between fermion and gauge field is dictated by the electric charge $e$ on the fermion, whereas $\xi$ is the covariant gauge fixing parameter. The corresponding action is 
\begin{equation}
    \mathcal{S} = \int d^dx \mathcal{L}\,,
\end{equation}
where the superscript $d$ denotes the number of spacetime dimensions in which the fermion and gauge fields are defined. For the standard QED, $d=4$, hereafter denoted as $\text{QED}_4$. A particularly interesting case arises when an abelian $\text{U}\left(1\right)$ gauge field propagates in $d_\gamma$ spacetime dimensions, while the fermion field is confined to a lower-dimensional spacetime of dimension $d_e$, with $d_e < d_\gamma$. This setup is referred to as reduced QED, denoted by $\text{RQED}_{d_\gamma,d_e}$.

In the literature, particular attention ha been given to $\text{RQED}_{4,3}$. This effective field theory is relevant to those condensed-matter systems whose low-energy excitations are gapless, exhibiting a relativistic, linear dispersion relation. In these systems, electrons are confined to a two-dimensional plane
$(d_e = 2+1)$ while the photons mediating their interactions have the liberty to propagate in a three-dimensional bulk $(d_\gamma = 3+1)$. Prominent examples include graphene and several artificial graphene-like materials~\cite{Wallace:1947qeg,Semenoff:1984dq,Novoselov:2005kj,Polini:2013gaa}. 
Just like QED$_4$ for massless fermions, the $\text{RQED}_{4,3}$ is multiplicatively renormalizable,~\cite{Albino:2022efn}. 
Till the two-loops level, $\text{RQED}_{4,3}$, $\text{RQED}_{3,2}$ and the general case of $\text{RQED}_{d_\gamma,d_e}$ have been 
studied in Refs.~\cite{Kotikov:2013eha,Teber:2014hna,Teber:2012de}.

In this article, we study the fermion propagator and its behavior under gauge transformations. It plays a central role in the computation of numerous physical observables of interest, including scattering cross sections, decay rates, and electromagnetic elastic and transition form factors.
Although Green functions are, in general, gauge dependent, this dependence cancels out in the final evaluation of physical observables, which are strictly gauge invariant. This cancellation can be systematically traced and verified within perturbation theory at every order of approximation.
However, in a nonperturbative setting, the use of inadequate truncation schemes in the underlying field-theoretic equations may lead to a loss of gauge invariance. These gauge transformations provide a powerful framework not only to restore gauge invariance of nonperturbative observables but also provide stringent constraints on admissible truncation schemes, thereby ensuring consistency with the key requirements of gauge invariance.

These transformations imply that a given Green function evaluated in two different covariant gauges is connected by definite and remarkably simple transformation rules in coordinate space. These rules were first obtained for the electron and photon propagators and for the electron–photon vertex by Landau and Khalatnikov~\cite{Landau:1955zz} and independently by Fradkin~\cite{Fradkin:1955jr} using canonical methods, hence dubbed as the Landau-Khalatnikov-Fradkin (LKF) transformations. Johnson and Zumino later derived them using functional techniques~\cite{Johnson:1959zz,Zumino:1959wt}.
A different approach for their derivation, through the introduction of a Stueckelberg field, was adopted 
in Ref.~\cite{Sonoda:2000kn}. It was subsequently  generalized to the non-Abelian case of quantum chromodynamics in Ref.~\cite{DeMeerleer:2018txc}, earlier studied in Ref.~\cite{Aslam:2015nia}.

 The LKF transformations have been studied for several Abelian theories, including QED2~\cite{Nicasio:2023zec}, QED3~\cite{Burden:1993gy,Aitchison:1997ua,Bashir:1999bd,Bashir:2000rv,Bashir:2000ur,Bashir:2002sp,Bashir:2004yt,Bashir:2004rg,Bashir:2005wt,Bashir:2009fv,Kotikov:2023qdm,Gusynin:2020cra}, QED4~\cite{Curtis:1990zs,Bashir:1994az,Bashir:2002sp,Kizilersu:2009kg}, QED$d$~\cite{Bashir:2004hh,Jia:2016wyu}, scalar QED~\cite{Villanueva-Sandoval:2013opv,Ahmadiniaz:2015kfq,Fernandez-Rangel:2016zac,Kotikov:2019bqo}, and reduced QED~\cite{Ahmad:2016dsb,James:2019ctc,Albino:2022efn}. Several of these works also employ requirements of gauge covariance to constrain the nonperturbative structure of the electron--photon vertex in the massless theory, while others analyze how the fermion propagator transforms under changes of gauge and compare the results with available perturbative expansions. Nonperturbative applications further provide valuable insight into the gauge invariance of dynamical chiral symmetry breaking and the emergence of fermion mass.

 In Ref.~\cite{Ahmad:2016dsb}, the gauge covariance properties of the massless fermion propagator in $\text{RQED}_{4,3}$ were analyzed. The study starts from the tree-level propagator in the Landau gauge. Landau–Khalatnikov–Fradkin transformations are then applied nonperturbatively to obtain its analytic form in other gauges. The resulting expressions are expanded in the weak-coupling limit and compared with the perturbative results of Ref.~\cite{Kotikov:2016yrn}. 
In this work, we extend this analysis to the massive electron propagator at two-loop order. We begin with the propagator in a suitable reference gauge, guided by the constraints of perturbation theory and multiplicative renormalizability. We then derive its all-order form in an arbitrary covariant gauge.
From this, we extract the weak-coupling expansion up to two loops for both massless and massive electrons. Wherever possible, we benchmark our results against existing perturbative calculations.
We also consider a hypothetical scenario in which the coupling is strong enough to trigger dynamical chiral symmetry breaking and generate an electron mass, studied in detail in Ref.~\cite{Albino:2022efn}. Using a combined analytical and numerical approach, we apply LKF transformations to obtain the results in other gauges. We then demonstrate that both the chiral fermion condensate and the fermion pole mass are gauge-invariant quantities, as expected. 
 
 In Sect.~\ref{formalism}, we briefly review the defining features of $\text{RQED}_{d\gamma,d_e}$. Sect.~\ref{sec3} has been devoted to the formalism and application of the LKF transformation to the fermion propagator in RQED$_{4,3}$. In Sect.~\ref{LKF-PE}, we present the two-loop expansion of the LKF-transformed fermion propagator. Section~\ref{renormalization} employs multiplicative renormalizability to derive the renormalized fermion propagator up to two loops. In Sect.~\ref{DCSB}, we apply LKF transformations to the dynamically generated fermion propagator in the reference gauge and analyze the gauge dependence of the mass function, the wave-function renormalization, the chiral fermion condensate, and the fermion pole mass. Finally, in Sect.~\ref{conclusion}, we summarize our findings 

\section{Reduced QED}\label{formalism}

As stated before, 
in RQED, photons roam around in $d_\gamma$ dimensions (bulk) whereas the fermions
 are confined to $d_e$ dimensions (brane), where we take $d_e\leq d_\gamma$. The RQED for the massive fermions is described by the action \cite{Marino:1992xi, Teber:2012de,Teber:2018goo, Heydeman:2020ijz, Guzman:2023tkm}  
\begin{equation}
\mathcal{S}_{d_{\gamma},d_{e}}\left[A_{\mu_{\gamma}},\psi\left(d_{e}\right)\right]=\int d^{d_{\gamma}}x\mathcal{L}_{d_{\gamma},d_{e}}\,,\label{eq:RQED2}
\end{equation}
where the Lagrangian 
is given by\,:
\begin{eqnarray}
\mathcal{L}_{d_{\gamma},d_{e}}&=&\bar{\psi}\left(x\right)\left(i\gamma^{\mu_{e}}D_{\mu_{e}}-m\right)\psi\left(x\right)\delta^{\left(d_{\gamma}-d_{e}\right)}\left(x\right)\notag\\
&&-\frac{1}{4}F_{\mu_{\gamma}\nu_{\gamma}}F^{\mu_{\gamma}\nu_{\gamma}}-\frac{1}{2\xi}\left(\partial_{\mu_{\gamma}}A^{\mu_{\gamma}}\right)^{2} \,.\label{eq:RQED3}
\end{eqnarray}
 $D_{\mu_{e}}$ is the covariant derivative, $m$ is the current mass of the fermion, the indices $\mu_{e}=0,\ldots,\left(d_{e}-1\right)$ for the $d_{e}-$dimensional
brane and $\mu_{\gamma}=\nu_{\gamma}=0,\ldots,\left(d_{\gamma}-1\right)$ with $d_{\gamma}>d_{e}$. The Dirac delta function $\delta^{\left(d_{\gamma}-d_{e}\right)}\left(x\right)$ ensures that the fermion fields defined in $d_e$ dimensions are localized in the $d_\gamma$ dimensions. 

In momentum space, the free photon propagator along the bulk dimensions has the standard QED form $\textit{i.e., }$
\begin{equation}
D_{\mu_{\gamma}\nu_{\gamma}}\left(p\right)=-\frac{i}{p^{2}}\left(g_{\mu_{\gamma}\nu_{\gamma}}-\frac{p_{\mu_{\gamma}}p_{\nu_{\gamma}}}{p^{2}}\right)+\xi\frac{p_{\mu_{\gamma}}p_{\nu_{\gamma}}}{p^{4}}\,,\label{eq:RQED4}
\end{equation}
where $\xi$ is a covariant gauge parameter and $\xi=0$ corresponds to the Landau gauge. The first part is the transverse
one, whereas the second one is referred to as the longitudinal part. 
In reduced $d_e$-brane, the photon propagator acquires the form~\cite{Teber:2012de, Teber:2014hna}
\begin{equation}
\hspace{-2mm} D_{\mu_{e}\nu_{e}}\left(p\right)=D\left(p^{2}\right) \left[ \left(g_{\mu_{e}\nu_{e}}-\frac{p_{\mu_{e}}p_{\nu_{e}}}{p^{2}}\right)+\widetilde{\xi} \, \frac{p_{\mu_{e}}p_{\nu_{e}}}{p^{2}}\right] \,,\label{eq:RQED5}
\end{equation}
where
\begin{equation}
D\left(p^{2}\right)=\frac{i}{\left(4\pi\right)^{\varepsilon_{e}}}\frac{\Gamma\left(1-\varepsilon_{e}\right)}{\left(-p^{2}\right)^{1-\varepsilon_{e}}}\,,\label{eq:RQED6}
\end{equation}
$\varepsilon_{e}=\left(d_{\gamma}-d_{e}\right)/2$ and the gauge parameter modifies to $\widetilde{\xi}=\left(1-\varepsilon_{e}\right)\xi$. It is crucial to the derivation of the LKF transformations for the electron propagator as detailed in the next section.

\section{LKF Tranformations for the Fermion propagtor}\label{sec3}
In an arbitrary covariant gauge, characterized by the parameter $\xi$, the general form of the Euclidean space fermion propagator in $d_e$-dimensions can be written in the following equivalent forms\,: 
\begin{equation}
S_{d_e}\left(p;\xi\right)=A\left(p;\xi\right)+\frac{iB\left(p;\xi\right)}{\slashed{p}}=\frac{F\left(p;\xi\right)}{i\slashed{p}-\mathcal{M}\left(p;\xi\right)}\,,\label{eq:1}
\end{equation}
where $F\left(p;\xi\right)$ and $\mathcal{M}\left(p;\xi\right)$ are called the
wavefunction renormalization and the mass function, respectively. The Lorentz decomposition of the corresponding Green function in the coordinate space reads as\,:
\begin{equation}
S_{d_e}\left(x;\xi\right)=\slashed{x}X\left(x;\xi\right)+Y\left(x;\xi\right) \,.\label{eq:2}
\end{equation}
The momentum and position space functions are related to each other through the Fourier
transformation\,: 
\begin{eqnarray}
S_{d_e}\left(x;\xi\right)&=&\int\frac{d^{d_e}p}{\left(2\pi\right)^{d_e}}\text{e}^{-ip\cdot x}S_{d_e}(p;\xi)\,,\label{eq:3a}\\
S_{d_e}\left(p;\xi\right)&=&\int d^{d_e}x \, \text{e}^{ip\cdot x}S_{d_e}\left(x;\xi\right)\,.\label{eq:3b}
\end{eqnarray} 
Therefore,
\begin{equation}
\hspace{-1mm} \slashed{x}X\left(x;\xi\right)+Y\left(x;\xi\right)=\int\frac{d^{d_e}p}{\left(2\pi\right)^{d_e}}\text{e}^{-ip\cdot x}\frac{F\left(p;\xi\right)}{i\slashed{p}-\mathcal{M}\left(p;\xi\right)}\,,\label{eq:10}
\end{equation}
At the lowest order in perturbation theory in the leading log approximation,  $F(p;\xi_0) = 1$ and $\mathcal{M}(p;\xi_0) = m$,~\cite{Albino:2022efn}, in the particular gauge $\xi_0 = 1/3$. In turn, it implies
\begin{eqnarray}
X\left(x;\xi_0\right) &=&-\frac{i}{x^{2}}\int\frac{d^{d_e}p}{\left(2\pi\right)^{d_e}}\text{e}^{-ip\cdot x}\frac{p\cdot x}{p^{2}+m^{2}}\,,\label{eq:11a}\\
\nonumber \\
Y\left(x;\xi_0\right)&=&-m\int\frac{d^{d_e}p}{\left(2\pi\right)^{d_e}}\text{e}^{-ip\cdot x}\frac{1}{p^{2}+m^{2}}\,.\label{eq:11b}
\end{eqnarray}
Note that $d^{d_e}p=p^{d_e-1}dp\sin^{d_e-2}\theta \, d\theta \, \Omega_{d_e-2}$,
where the solid angle is given by $\Omega_{d_e-2}=2\pi^{(d_e-1)/2}/\Gamma\left((d_e-1)/2\right)$. We can now proceed to evaluate these Fourier integrals in three space-time dimensions and then apply the LKF transformations to find the corresponding expressions in an arbitrary covariant gauge $\xi$.
For $d_e=3$, after performing angular and momentum integration in Eqs.~(\ref{eq:11a}) and~(\ref{eq:11b}), the fermion Green's function at the tree level
 can be written as
\begin{align}
S_{d_e}\left(x;\xi_0\right) & =\slashed{x}X\left(x;\xi_0\right)+Y\left(x;\xi_0\right)\;\nonumber \\
 & =-\slashed{x}\frac{1}{4\pi}\left(\frac{1+mx}{x^{3}}\right)\text{e}^{-mx}-\frac{m}{4\pi x}\text{e}^{-mx}\,.\label{eq:RQ3}
\end{align}
Recall that the general form  of the LKF transformation  for the fermion propagator in $\text{RQED}_{d_\gamma,d_e}$\cite{Teber:2012de, Gorbar:2001qt, Ahmad:2016dsb} is\,:
\begin{equation}
S_{d_{e}}\left(x;\xi\right)=S_{d_{e}}\left(x;\xi_0\right)\text{e}^{-i\left[\widetilde{\Delta}_{d_{e}}\left(0;\varepsilon_{e}\right)-\widetilde{\Delta}_{d_{e}}\left(x;\varepsilon_{e}\right)\right]} \,,\label{eq:RQ4}
\end{equation}
with 
\begin{equation}
\widetilde{\Delta}_{d_{e}}\left(x;\varepsilon_{e}\right)=-if\left(\varepsilon_{e}\right)\xi e^{2}\mu^{4-d_{\gamma}}\int\frac{d^{d_{e}}q}{\left(2\pi\right)^{d_{e}}}\frac{\text{e}^{-iq\cdot x}}{q^{4-2\varepsilon_{e}}}\,.\label{eq:RQ5}
\end{equation}
The introduction of a new momentum scale $\mu$ guarantees that $e^2$ is dimensionless in arbitrary $d_\gamma$ dimensions and $f\left(\varepsilon_{e}\right)=\Gamma\left(2-\varepsilon_{e}\right)/\left(4\pi\right)^{\varepsilon_{e}}$. Integration over $q$ leads to
\begin{equation}
\widetilde{\Delta}d_{e}\left(x;\varepsilon_{e}\right) = if\left(\varepsilon_{e}\right)\xi e^{2}\frac{\Gamma\left(\frac{d_{e}-a}{2}\right)}{2^{a}\pi^{d_{e}/2}\Gamma\left(\frac{a}{2}\right)}\left(\mu x\right)^{a-d_{e}}\,,\label{eq:RQ6}
\end{equation}
where $a=4-2\varepsilon_{e}$ and $d_{\gamma}=2\varepsilon_{e}+d_{e}$. This matches with the result in \cite{Ahmad:2016dsb}, and setting $d_\gamma = d_e = d$, we can retrieve the LKF transformation for standard QED. Therefore, in the case of a graphene–like material, for which $d_{e}=3$ and $d_{\gamma}=4$, it follows that $\varepsilon_{e}=1/2$. Consequently\,:
\begin{eqnarray}
S_{3}\left(x;\xi\right)&=&S_{3}\left(x;\xi_0\right)\text{e}^{-i\left[\widetilde{\Delta}_{3}\left(0;1/2\right)-\widetilde{\Delta}_{3}\left(x;1/2\right)\right]}  \,, \label{FP-Graphene} \\ [2mm]
\widetilde{\Delta}_{3}\left(x;1/2\right)  &=&-\frac{i}{4}\left(\xi-\xi_0\right)e^{2}\frac{\Gamma\left(\varepsilon\right)}{2^{3}\pi^{3/2}\Gamma\left(\frac{3}{2}\right)}\left(\mu x\right)^{-2\varepsilon} \,,\nonumber \\
&=&  -i \nu \left[\frac{1}{\varepsilon}-\gamma-2\ln\left(\mu x\right)+\mathcal{O}\left(\varepsilon\right)\right] \,,
\label{eq:RQ7} 
\end{eqnarray}
where we have conveniently defined 
$\varepsilon = \varepsilon_e -1/2$ to adopt a compact notation. We have also carried out a Taylor expansion around $\varepsilon = 0$. At $x=0$, we regularize this expression by introducing a cut-off $x_{\text{min}}$, such that
\begin{equation}
\widetilde{\Delta}_{3}\left(x;1/2\right)-\widetilde{\Delta}_{3}\left(x_{\text{min}};1/2\right)=i\ln\left(\frac{x}{x_{\text{min}}}\right)^{2\nu} \,,\label{eq:RQ9}
\end{equation}
and $\nu$, introduced in Eq.~(\ref{eq:RQ7}), is defined as
\begin{eqnarray}
\nu\equiv  \frac{\left(\xi-\xi_0\right)\alpha}{4\pi} \,.
\end{eqnarray}
Substituting Eq.~(\ref{eq:RQ9}) in Eq.~(\ref{FP-Graphene}) and then making use of Eq.~(\ref{eq:RQ3}) yields the following final expression for the LKF transformed fermion propagator in the coordinate space in an arbitrary covariant gauge $\xi$\,:
\begin{equation}
\hspace{-3mm} S_{d_e}\left(x;\xi\right)= - \frac{\text{e}^{-mx}}{4\pi x}\left[\slashed{x}\left(\frac{1+mx}{x^{2}}\right) + m \right] \left(\frac{x}{x_{\text{min}}}\right)^{-2\nu} \hspace{-4mm} .\label{eq:RQ11}
\end{equation}
Transforming these expressions back to the momentum space, though tedious, is straightforward: performing the inverse Fourier transform, we obtain
\begin{eqnarray}
 && \hspace{-11mm} A\left(p;\xi\right) \hspace{-1mm}
  = \hspace{-0.8mm} - \frac{\left(mx_{\text{min}}\right)^{2\nu}}{p} \Gamma\left(1-2\nu\right) \left(1+\frac{p^{2}}{m^{2}}\right)^{\nu-\frac{1}{2}} \hspace{-2mm} \sin \theta_1 ,\label{eq:RQ-16}\\
&& \hspace{-11mm}  B\left(p;\xi\right) \hspace{-1mm} = 
\hspace{-0.8mm} -\frac{m \mathcal{D}}{p}\left(mx_{\text{min}}\right)^{2\nu} \hspace{-0.8mm}  \Gamma\left(-1-2\nu\right) \hspace{-1mm} \left(1+\frac{p^{2}}{m^{2}}\right)^{\nu-\frac{1}{2}}  \hspace{-6mm} ,
\label{eq:RQ-16b}
\end{eqnarray}
where 
\begin{eqnarray}
&& \hspace{-1.05cm} \theta_1 = 
\left(1 - 2\nu \right) \tan^{-1}\left(\frac{p}{m}\right) \,,  \\
&& \hspace{-1.05cm} \theta_2 = 
2\nu\tan^{-1}\left(\frac{p}{m}\right) \,,  \\ 
&& \hspace{-1cm} \mathcal{D} =\left(1+2\nu\right)\sqrt{1+\frac{p^{2}}{m^{2}}} \sin \theta_2 \notag\\
&& \hspace{-6mm} +\left[\left(1+2\nu\frac{p^{2}}{m^{2}}\right)\sin \theta_1 
-\frac{p}{m} \left(1 + 4\nu^{2}\right)\cos \theta_1 \right]\,.
\label{eq:RQ-21a}
\end{eqnarray} 
 One can readily express the more familiar mass function, $\mathcal{M}\left(p;\xi\right)$, and the fermion wavefunction renormalization, $\mathcal{F}\left(p;\xi\right)$, in terms of $A\left(p;\xi\right)$ and $B\left(p;\xi\right)$ through the use of Eq.~(\ref{eq:1}). These expressions are listed below\,:
\newpage
\begin{eqnarray}
&& \hspace{-1.4cm} \mathcal{M}\left(p;\xi\right)= m \, \frac{1}{\mathcal{D}} \frac{p^{2}}{m^{2}}  \, 2 \nu \left(1 + 2 \nu \right) \, \sin \theta_1 \,,\label{Mfun1a}\\ 
&& \hspace{-1.22cm}\mathcal{F}\left(p;\xi\right)=\frac{m}{p} \frac{1}{\mathcal{D}} \left(mx_{\text{min}}\right)^{2\nu}\left(1+\frac{p^{2}}{m^{2}}\right)^{\nu-\frac{1}{2}}\notag\\
&& \hspace{-6mm} \times \Gamma\left(-2\nu-1\right) \left[\mathcal{D}^{2} +4 \frac{p^{2}}{m^{2}} \, \nu^2 (1+ \nu^2) \sin^{2} \theta_1 \right]\,. \label{ffunc1a}
\end{eqnarray}
This section completes the discussion on a general 
nonperturbative expression for the LKF transformed fermion propagator in an arbitrary covariant gauge $\xi$, expressed within the definition of $\nu=\alpha(\xi-\xi_0)/4 \pi$, where $\xi_0$ defines the starting gauge where we take the fermion propagator to be its bare counterpart. 

\section{Perturbative Expansion}
\label{LKF-PE}

We reiterate that the explicit expressions for the functions $A\left(p;\xi\right)$ and $B\left(p;\xi\right)$, displayed in Eqs.~(\ref{eq:RQ-16}) and~(\ref{eq:RQ-16b}), or alternatively, for $\mathcal{M}\left(p;\xi\right)$ and 
$\mathcal{F}\left(p;\xi\right)$, as derived in Eqs.~(\ref{Mfun1a}) and~(\ref{ffunc1a}),
are nonperturbative in nature. However, in this sub-section, we expand these results in powers of $\alpha$ or $\nu$. Such an expansion is customarily carried out through Feynman diagrams in the more familiar formalism of S-matrix perturbation theory. 
Subsequently, we perform further sub-expansions in the limiting kinematic case of interest: $p^2 \gg m^2$. On the one hand, it enables an analysis of the ultraviolet behavior of the fermion propagator; on the other hand, it facilitates a direct comparison with the massless case previously studied in the literature by explicitly setting the fermion mass $m=0$. 
We start by carrying out a perturbative expansion of the functions $A(p;\xi)$ and $B(p;\xi)$, defined in Eqs.~(\ref{eq:RQ-16}) and (\ref{eq:RQ-16b}), in the parameter
$\alpha$ or $\nu$ till two-loop order, i.e., ${\cal O}(\alpha^2)={\cal O}(\nu^2)$.
Within this series, we also perform a small-mass expansion in powers of
$(m^2/p^2)$:

\begin{widetext}
    \begin{eqnarray}
A(p;\xi) & =&-\frac{m}{p^{2}}\Biggl[1-\frac{m^{2}}{p^{2}}+\nu\left\{ 2\gamma+\log\left(p^{2}x_{\text{min}}^{2}\right) - \frac{m}{p} \, \pi +\frac{m^{2}}{p^{2}} \left(3-2\gamma-\log\left(p^{2}x_{\text{min}}^{2}\right)\right)\right\}\notag\\
 &&+\nu^{2}\Bigg\{2\gamma^{2}-\frac{\pi^{2}}{6}+2\gamma\log\left(p^{2}x_{\text{min}}^{2}\right)+\frac{1}{2}\log^{2}\left(p^{2}x_{\text{min}}^{2}\right)+\frac{m}{p} \, \pi \left(2 -2 \gamma-\log\left(p^{2}x_{\text{min}}^{2}\right)\right)\notag\\
 && \hspace{8mm} + \frac{m^{2}}{p^{2}} \left(-2+6\gamma-2\gamma^{2}+\frac{\pi^{2}}{6}+ \left( 3 - 2 \gamma \right) \log\left(p^{2}x_{\text{min}}^{2}\right)-\frac{1}{2}\log^{2}\left(p^{2}x_{\text{min}}^{2}\right)\right)\Bigg\}\Biggr]\,,\label{AfuncC2} \\
B(p;\xi) & =&-1+\frac{m^{2}}{p^{2}} +\nu\left[\left(2-2\gamma-\log\left(p^{2}x_{\text{min}}^{2}\right)\right)+\frac{m^{2}}{p^{2}}\left(-1+2\gamma+\log\left(p^{2}x_{\text{min}}^{2}\right)\right)\right]\notag\\
 &&+\nu^{2}\Bigg[\left(-4+4\gamma-2\gamma^{2}+\frac{\pi^{2}}{6}+2 \left(1 - \gamma\right) \log\left(p^{2}x_{\text{min}}^{2}\right)-\frac{1}{2}\log^{2}\left(p^{2}x_{\text{min}}^{2}\right)\right)\notag\\
 && \hspace{8mm} +\frac{m^{2}}{p^{2}}\left(-2\gamma+2\gamma^{2}-\frac{\pi^{2}}{6}- \left(1 - 2 \gamma \right) \log\left(p^{2}x_{\text{min}}^{2}\right)+\frac{1}{2}\log^{2}\left(p^{2}x_{\text{min}}^{2}\right)\right)\Bigg]\,.\label{BfuncC2}
\end{eqnarray}
\end{widetext}
Using Eqs.~(\ref{eq:1}),~(\ref{AfuncC2}), and~(\ref{BfuncC2}), the mass function and the fermion wavefunction renormalization, to  leading order in $m^2/p^2$ and the second order in $\mathcal{O}\left(\nu^2\right)$ in RQED, take the form\,:
\begin{widetext}
\begin{eqnarray}
&& \hspace{-6.5mm} \mathcal{M}(p;\xi) = m\left[1+ \nu \left\{ 2-\pi\left(\frac{m}{p}\right)+4\left(\frac{m^{2}}{p^{2}}\right)\right\}
+4 \, \nu^2\left(\frac{m^{2}}{p^{2}}\right)\right]  \,, \label{MfuncC2}\\
&& \hspace{-5mm} \mathcal{F}(p;\xi) = 1+ \nu 
\left[ -2(1 - \gamma)  +\log\left(p^{2}x_{\text{min}}^{2}\right)+3\left(\frac{m^{2}}{p^{2}}\right)\right] \nonumber \\
&& \hspace{7mm}
+ \nu^2\biggl[2\gamma^{2}-\frac{\pi^{2}}{6} + 2 (1 -\gamma)  \left\{2 - \log\left(p^{2}x_{\text{min}}^{2}\right) \right\} +\frac{1}{2}\log^{2}\left(p^{2}x_{\text{min}}^{2}\right)
 + 3 \left(\frac{m^{2}}{p^{2}}\right)\left\{2\gamma+\log\left(p^{2}x_{\text{min}}^{2}\right)\right\}\biggr] \,. \label{FfuncC2}
\end{eqnarray}
In the massless limit, i.e., $m \to 0$, Eqs.~(\ref{MfuncC2}) and~(\ref{FfuncC2}) reduce to\,:
\begin{eqnarray}
&& \hspace{-7.5mm} \mathcal{M}(p;\xi) =0 \,, \\
&& \hspace{-6mm} \mathcal{F}(p;\xi)  =1+ \nu \left[-2(1-\gamma)+\log\left(\frac{p^{2}}{\Lambda^{2}}\right)\right]+ \nu^2\left[2\gamma^{2}-\frac{\pi^{2}}{6} + 2 (1 -\gamma)  \left\{ 2 - \log\left(\frac{p^{2}}{\Lambda^{2}}\right) \right\} +\frac{1}{2}\log^{2}\left(\frac{p^{2}}{\Lambda^{2}}\right) \right] \,,
\label{FPropc1}
\end{eqnarray}
\end{widetext}
where we have made the identification $x_{\text{min}}\equiv1/\Lambda$. We find our results to be in full agreement with Refs.~\cite{Ahmad:2016dsb, Kotikov:2013eha}, thereby confirming the massless limit of our calculation. Note that the leading logarithms expression for $\mathcal{F}(p;\xi)$ becomes
\begin{eqnarray}
 \mathcal{F}(p;\xi)  =1+ \nu \log\left(\frac{p^{2}}{\Lambda^{2}}\right)  +\frac{\nu^2}{2!} \log^{2}\left(\frac{p^{2}}{\Lambda^{2}}\right) + \cdots \,.
\label{FLL}
\end{eqnarray}
Recalling that $\nu = \alpha \left(\xi - \xi_0\right)/{4 \pi}$ and then comparing this expression with the direct computation of the wavefunction renormalization in perturbation theory,~\cite{Albino:2022efn,Guzman:2023tkm}, {\em i.e.,}
\begin{eqnarray}
 \mathcal{F}(p;\xi)  =1+ \frac{\alpha \left(\xi - 1/3\right)}{4 \pi} \log\left(\frac{p^{2}}{\Lambda^{2}}\right)  + \cdots \,,
\label{FPT}
\end{eqnarray}
we straightforwardly conclude that $\xi_0 = 1/3$. 
Moreover, the series in Eq.~(\ref{FLL}) suggests a readily verifiable multiplicatively renormalizable expression\,:
\begin{eqnarray}
 \mathcal{F}(p;\xi) = \left(  \frac{p^2}{\Lambda^2} \right)^{\nu} \,.
\label{FMR}
\end{eqnarray}
Note that this expression for the wavefunction renormalization coincides with that of massless QED$_4$~\cite{Curtis:1990zs}, with the sole distinction that $\xi_0 = 0$ in QED$_4$, whereas $\xi_0 = 1/3$ in RQED. Consequently, while the Landau gauge provides a natural starting point for implementing LKF transformations in QED$_4$, the corresponding choice in RQED is $\xi_0 = 1/3$.

In this perturbative regime, the behavior of the mass function $\mathcal{M}(p;\xi)$ highlights a notable aspect of perturbation theory: it vanishes identically as $m \to 0$, regardless of the order considered. This stands in sharp contrast to the dynamical breaking of chiral symmetry, which can generate a nonzero mass even in the chiral limit when the coupling $\alpha$ is sufficiently large. The nonperturbative aspects of the theory will be examined later in the article.
Prior to that study, we address the renormalization of ultraviolet divergences in the following section. \vspace{3mm}

\section{Renormalization}\label{renormalization}

Fermion propagator is logarithmically divergent and needs to be renormalized. As the mass function $\mathcal{M}(p;\xi)$ does not exhibit these divergences, we focus our attention on the  wavefunction renormalization $\mathcal{F}(p;\xi)$ and on its renormalization 
at the two-loop order in the RQED. 
\begin{eqnarray} 
\frac{F_{R}\left(m^2/p^2,p^2/\mu^2;\xi\right)}{\mathcal{F}\left(m^2/p^2,p^2/\Lambda^2;\xi\right)} = \mathcal{Z}^{-1}_2\left(\mu^2;\Lambda^2\right)\,,\label{eq:67} \\ \nonumber
\end{eqnarray}
where we have taken the liberty to adjust the notation again in a straightforward and convenient manner.
Till the two-loops order, {\em i.e.,} ${\cal {O}}(\nu^2)={\cal {O}}(\alpha^2)$, the renormalization constant can be expanded out as follows\,: 
\begin{eqnarray}
&&  \hspace{-1.1cm} \mathcal{Z}^{-1}_2\left(\mu^2;\Lambda^2\right) = 1+\nu\mathcal{Z}^{-1}_{2;1}+\nu^2\mathcal{Z}^{-1}_{2;2} \,, \nonumber \\
&& \hspace{1cm}  = 1-\nu\log\left(\frac{\mu^{2}}{\Lambda^{2}}\right)
+\frac{1}{2}\nu^2\log^{2}\left(\frac{\mu^{2}}{\Lambda^{2}}\right)\,. 
\label{eq:mlZ2}
\end{eqnarray}
We can now proceed to compute the wavefunction renormalization for massive fermions till ${\cal O}(\alpha^2)$:
\begin{widetext}
\begin{eqnarray}
&&  \hspace{-1cm} F_{R}\left(m^2/p^2,p^2/\mu^2;\xi\right)  =  1+\nu\left[2\left(\gamma-1\right)+\log\left(\frac{p^2}{\Lambda^2}\right)+3\left(\frac{m^{2}}{p^{2}}\right)+\mathcal{Z}^{-1}_{2;1}\right]+\nu^2\left[\frac{1}{2}\log^2\left(\frac{p^2}{\Lambda^2}\right) \right.\nonumber \\
 &&  \hspace{-1cm} \left.+2\gamma\log\left(\frac{p^2}{\Lambda^2}\right)+2\gamma^{2}-\frac{\pi^{2}}{6}+\left\{ 2\left(\gamma-1\right)+\log\left(\frac{p^2}{\Lambda^2}\right)\right\}\left(\mathcal{Z}^{-1}_{2;1}-2\right)+\mathcal{Z}^{-1}_{2;2}
 +3\left(\frac{m^{2}}{p^{2}}\right)\left\{ 2\gamma+\log\left(\frac{p^2}{\Lambda^2}\right)+\mathcal{Z}^{-1}_{2;1}\right\}\right] \,.
 \label{eq:RQ-27}
\end{eqnarray}
Substituting the values of the renormalization constant ${\cal Z}_2$, we can write
\begin{eqnarray}
&&  \hspace{-1.2cm} F_{R}\left(m^2/p^2,p^2/\mu^2;\xi\right)  =  1+\nu\left[2\left(\gamma-1\right)+\log\left(\frac{p^2}{\mu^2}\right)+3\left(\frac{m^{2}}{p^{2}}\right)\right] \nonumber \\
 && \hspace{8mm} +\nu^2\left[\frac{1}{2}\log^2\left(\frac{p^2}{\mu^2} \right) +2\left(\gamma-1\right)\log\left(\frac{p^2}{\mu^2}\right)+2\gamma^{2}-\frac{\pi^{2}}{6}-4\left(\gamma-1\right)+3\left(\frac{m^{2}}{p^{2}}\right)\left\{ 2\gamma+\log\left(\frac{p^2}{\mu^2}\right)\right\}\right]\,.
\end{eqnarray}
In the massless limit,
\begin{eqnarray}
\hspace{-2mm} F_{R}\left(p^2/\mu^2;\xi\right)  =  1+\nu\left[2\left(\gamma-1\right)+\log\left(\frac{p^2}{\mu^2}\right)\right]+\nu^2\left[\frac{1}{2}\log^2\left(\frac{p^2}{\mu^2}\right) 
 +2\left(\gamma-1\right)\log\left(\frac{p^2}{\mu^2}\right)+2\gamma^{2}-\frac{\pi^{2}}{6}-4\left(\gamma-1\right)\right] ,
\end{eqnarray}
\end{widetext}
which agrees with the massless two loop perturbative result, Ref.~\cite{Kotikov:2013eha}. Naturally, the leading logarithmic expression is\,:
\begin{eqnarray}
F_{R}\left(p^2/\mu^2;\xi\right) = 
1+ \nu \log\left(\frac{p^{2}}{\mu^{2}}\right)  +\frac{\nu^2}{2!} \log^{2}\left(\frac{p^{2}}{\mu^{2}}\right) + \cdots \,,\nonumber
\end{eqnarray}
which leads to the multiplicatively renormalized massless fermion propagator\,:
\begin{eqnarray}
F_{R}\left(p^2/\mu^2;\xi\right) = 
 \left(  \frac{p^2}{\mu^2} \right)^{\nu} \,.
\end{eqnarray}
In the next section, we study the fermion propagator which is generated through dynamical chiral symmetry breaking, its gauge covariance properties and the gauge invariance of related physical observables.

\section{Gauge Covariance and dynamically Generated Masses}\label{DCSB}
In this section we present the results obtained by applying the LKF
transformation to the fermion propagator obtained by solving the gap
equation using different Ansätze for the fermion-photon vertex above
critical coupling, where fermion mass is generated through dynamical
chiral symmetry breaking. Specifically, we consider the bare vertex,
the Ball-Chiu (BC) vertex \cite{Ball:1980ay}, and the Curtis-Pennington (CP) vertex \cite{Curtis:1993py}.
The purpose of this analysis is to examine the gauge covariance properties
of the fermion propagator and to compare the behavior of physical
observables obtained directly from solving the gap equation in different
covariant gauges with those obtained through LKF transformation. 

To apply LKF transformation, we start from a nonperturbative solution
of the fermion propagator with an initial choice of $\xi=\xi_{0}$.
Restricting ourselves to the physically relevant case $d_{e}=3$,
the coordinate space scalar functions $X(x;\xi_{0})$ and $Y(x;\xi_{0})$
can be obtained through the Fourier transformation of the momentum-space
propagator. Explicitly, on computing the angular integrals, one finds
\begin{align}
X(x;\xi_0) &= \frac{1}{2\pi^{2}x^{2}}
\int dp\, p^{2}\,
\frac{F(p;\xi_0)}
{p^{2}+\mathcal{M}^{2}(p;\xi_0)}
\nonumber\\
&\hspace{1.5cm}\times
\left[
\cos(px)-\frac{\sin(px)}{px}
\right],
\label{DM1a}
\\[2mm]
Y(x;\xi_0) &= -\frac{1}{2\pi^{2}}
\int dp\, p^{2}\,
\frac{F(p;\xi_0)\mathcal{M}(p;\xi_0)}
{p^{2}+\mathcal{M}^{2}(p;\xi_0)}
\frac{\sin(px)}{px}.
\label{DM1b}
\end{align}
These expressions encode the full nonperturbative information contained within the wavefunction renormalization $F(p;\xi_0)$ and the dynamically generated mass function $\mathcal{M}(p;\xi_0)$. The latter emerges through the dynamical breakdown of chiral symmetry when the interaction strength exceeds a critical value. The functions $X(x;\xi_0)$ and $Y(x;\xi_0)$ therefore provide the starting point for implementing the LKF transformations in coordinate space. For $d_\gamma=4$ and the initially massless fermions, it takes a particularly simple multiplicative form\,:
\begin{align}
X(x;\xi) &= X(x;\xi_0)
\left(
\frac{x}{x_{\rm min}}
\right)^{-2\nu},
\label{DM2a}
\\
Y(x;\xi) &= Y(x;\xi_0)
\left(
\frac{x}{x_{\rm min}}
\right)^{-2\nu}.
\label{DM2b}
\end{align}
These relations demonstrate that the gauge dependence of the massless fermion propagator is entirely determined by a multiplicative power-law factor in coordinate space. Consequently, once the propagator is known in a reference gauge, its form in any other covariant gauge can be reconstructed exactly through LKF transformation.

\begin{figure}[t]
\begin{center}
\includegraphics[width=0.95\columnwidth]{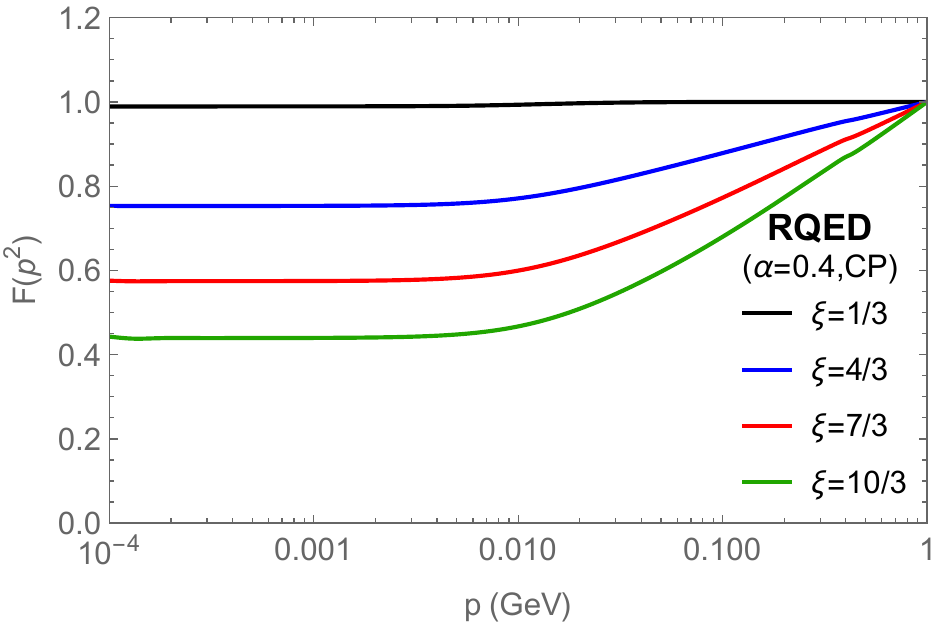}

\vspace{2mm}

\includegraphics[width=0.95\columnwidth]{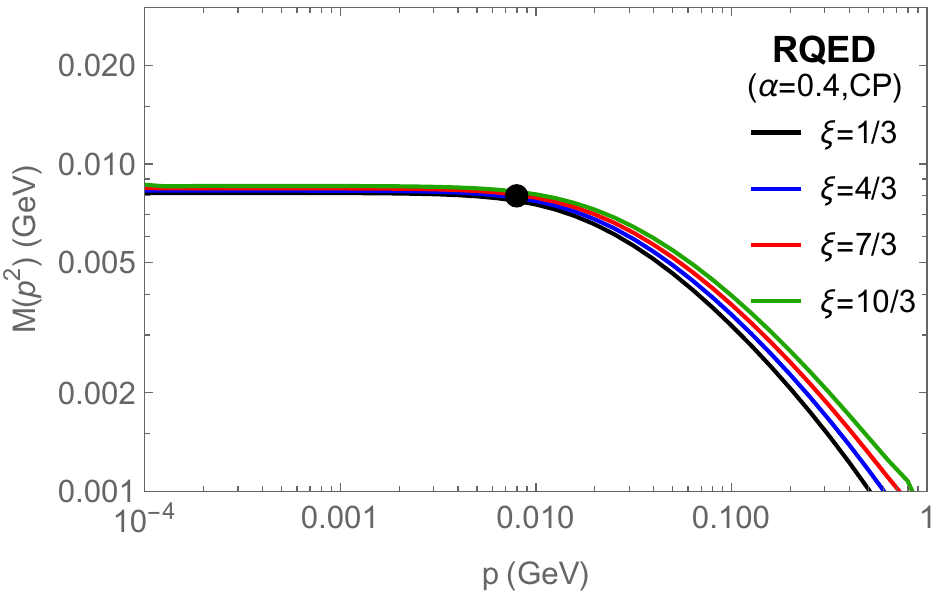}
\end{center}
\caption{
Wavefunction renormalization (top panel) and dynamically generated mass function (bottom panel) for several values of the covariant gauge parameter $\xi$. The nearly constant black region in the lower panel corresponds to the Euclidean pole mass, illustrating its gauge independence.
}
\label{Fig1}
\end{figure}

\begin{figure}[t]
\begin{center}
\includegraphics[width=0.95\columnwidth]{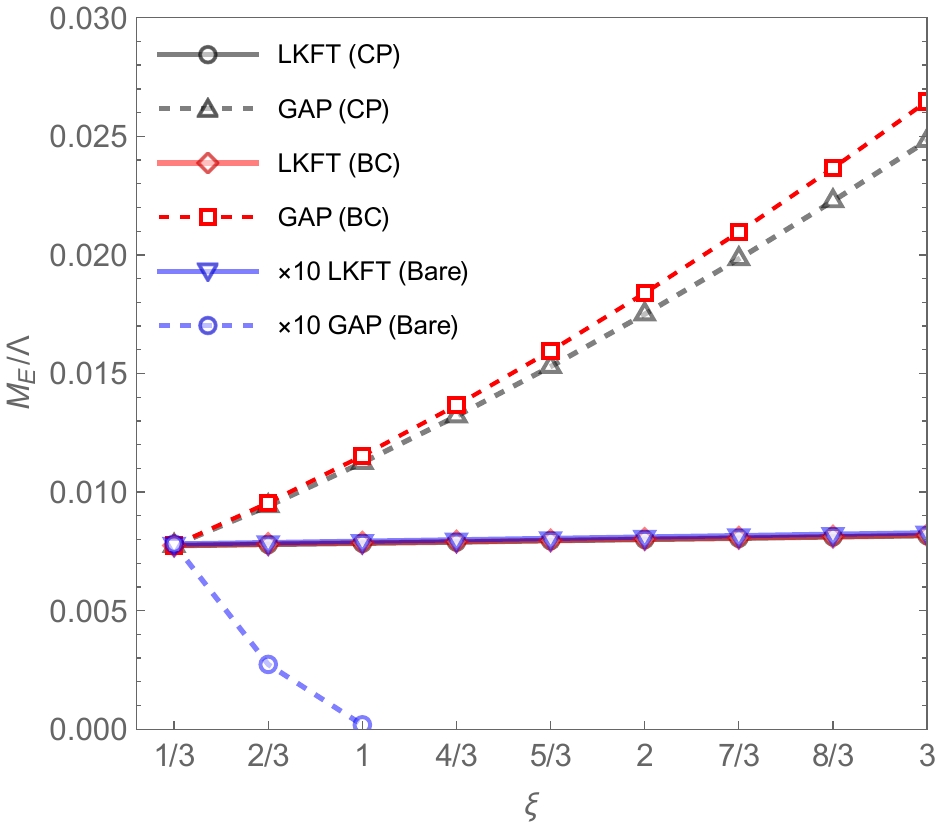}

\vspace{2mm}

\includegraphics[width=0.95\columnwidth]{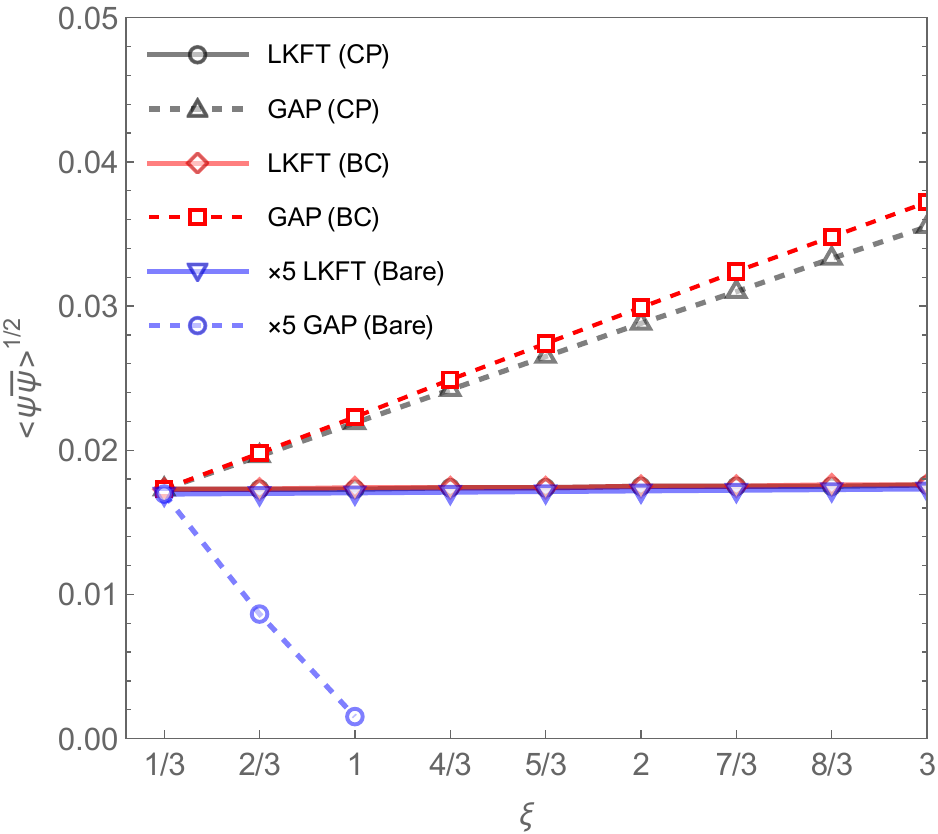}
\end{center}
\caption{
Euclidean pole mass (top panel) and chiral condensate (bottom panel) for several values of the covariant gauge parameter $\xi$. Solid and dashed curves represent 
the values obtained from LKF-transformed propagators and independently solving the gap equation, respectively. For clarity of presentation, the pole mass and chiral condensate results corresponding to the bare vertex are rescaled by factors of 10 and 5, respectively.
}
\label{Fig2}
\end{figure}

To recover the momentum-space propagator in an arbitrary gauge, we perform the inverse Fourier transformation. For $d_e=3$, this yields
\begin{eqnarray}
&& \hspace{-8mm} A(p;\xi) = -\frac{2}{\pi}
\int dk\, k^{2}\,
\frac{
F(k;\xi_0)\mathcal{M}(k;\xi_0)
}
{
k^{2}+\mathcal{M}^{2}(k;\xi_0)
}
f_A(p,k,\nu) \,,
\label{DM3a}
\\  [2mm]
&& \hspace{-8mm} B(p;\xi) = -\frac{2}{\pi}
\int dk\, k^{2}\,
\frac{
F(k;\xi_0)
}
{
k^{2}+\mathcal{M}^{2}(k;\xi_0)
}
f_B(p,k,\nu) \,,
\label{DM3b}
\end{eqnarray}
where the kernels $f_A\equiv f_A(p,k,\nu)$ and $f_B\equiv f_B(p,k,\nu)$ are defined as
\begin{eqnarray}
&& \hspace{-8mm} f_A
 = 
\frac{1}{pk}
\int  dx\,
\sin(px)\sin(kx)
\left(
\frac{x}{x_{\rm min}}
\right)^{-2\nu} \nonumber
\\ [2mm]
&& \hspace{-2.5mm} =
\frac{\nu\,\Gamma(-2\nu)\sin(\pi\nu)}
{pk \left({x_{\rm min}}\right)^{-2\nu}} 
\left[
(k+p)^{2\nu-1}  -   |k-p|^{2\nu-1}
\right] , \nonumber
\end{eqnarray}
and
\begin{eqnarray}
&& \hspace{-4mm}  f_B
\hspace{-1mm} = \hspace{-2mm}
\int \hspace{-1mm} dx
\hspace{-1mm} \left[
\cos(px) \hspace{-1mm} - \hspace{-1mm} \frac{\sin(px)}{px}
\right] \hspace{-1mm} \left[
\cos(kx) \hspace{-1mm} - \hspace{-1mm}\frac{\sin(kx)}{kx}
\right] \hspace{-1mm} 
\left( \hspace{-0.8mm}
\frac{x}{x_{\rm min}}
 \hspace{-0.8mm} \right)^{\hspace{-1mm}  -2\nu} \hspace{-3mm} ,
\nonumber
\\[2mm]
&& \hspace{-4mm} =
\frac{\nu\,\Gamma(-2\nu-1)\sin(\pi\nu)}
{pk \left({x_{\rm min}}\right)^{-2\nu} } \hspace{-1mm}
\left[
\lambda_{-}|k-p|^{2\nu-1}
\hspace{-1mm} - \hspace{-1mm} \lambda_{+}(k+p)^{2\nu-1}
\right],
\nonumber
\end{eqnarray}
with
\begin{equation}
\lambda_{\pm}
=
k^{2}+p^{2}
\pm kp(1-2\nu) \,. \nonumber
\end{equation}
The kernels contain integrable singularities at $k=p$, arising from the terms proportional to $|k-p|^{2\nu-1}$. These singularities are harmless and can be isolated analytically by separating a small region around $k=p$. Indeed,
\begin{equation}
\int_{p-\epsilon}^{p+\epsilon}
dk\,
|k-p|^{2\nu-1}
=
\frac{\epsilon^{2\nu}}{\nu} \,,
\end{equation}
which remains finite for positive $\nu$. After removing this integrable contribution, the remaining momentum integrations can be evaluated numerically in a stable manner. We apply LKF transformation on the fermion
propagator with the initial choice $\xi_{0}=1/3$ for which $F(p,1/3)\approx1$ \cite{Albino:2022efn}
for all three vertices considered in this work. The resulting wavefunction
renormalization and mass functions for several values of the gauge
parameter $\xi$ are shown in Fig.~\ref{Fig1} for the CP vertex. As expected, both
the functions exhibit explicit gauge dependence. Nevertheless, the Euclidean
pole mass remains essentially unchanged under variations of the gauge
parameter. This behavior is observed for all three vertices, as shown
in Fig.~\ref{Fig2}, where we compare the gauge dependence of pole mass and chiral
condensate obtained from the LKF-transformed propagators with those
extracted from independently solving the gap equation for each value
of $\xi$ under the same truncation scheme. The plots clearly show that the values of the observables
obtained through the LKF transformation remain constant
over the range of gauge parameter. In contrast, when the
gap equation is solved independently for different values $\xi$, the observables
exhibit a significant gauge dependence, particularly in the case of
the bare vertex. This result strongly suggests that the
LKF transformation preserves the essential gauge-covariant structure
of the fermion propagator more faithfully than the direct solutions
of the truncated gap equation.

There is an intriguing feature that appears in the comparison between
the results of BC and CP vertices for the initial choice of gauge parameter
$\xi_{0}=1/3$ for which $F\approx1$. We find that the corresponding
mass functions for both vertices are almost the same. Consequently, the
values of pole masses and chiral condensates extracted from the LKF transformed
fermion propagator also agree. This is despite the fact that direct
numerical solutions of the gap equation using the BC and CP vertices
display substantial differences in these quantities when solved independently
in different gauges. The near equivalence of the BC and CP based LKF
results indicates, that once the fermion propagator satisfies $F\approx1$,
the transverse structure distinguishing the CP vertex from the BC
contributes only weakly to the gauge-covariant evolution encoded by
the LKF transformation. In other words, the dominant gauge dependence
appears to be governed primarily by the longitudinal part of the vertex
which is constrained by the Ward-Green-Takahashi identity, while the
additional transverse terms mainly affect the dynamics of gap equation
solutions rather than the LKF mapping itself.


\newpage
\section{Conclusions}\label{conclusion}

In this article, we have investigated the gauge covariance properties of the fermion propagator in RQED$_{4,3}$ through the framework of Landau--Khalatnikov--Fradkin transformations. Starting from the fermion propagator in a suitable reference gauge, we derived analytical nonperturbative expressions for the propagator in an arbitrary covariant gauge, valid for both massive and massless fermions.

The perturbative expansion of these expressions up to two-loop order reproduces the known results available in the literature for the massless theory and extends them to the massive case. In particular, the comparison with perturbation theory and the requirements of multiplicative renormalizability identify $\xi_0=1/3$ as the natural reference gauge in RQED$_{4,3}$. In this gauge, the leading logarithmic contribution to the wavefunction renormalization vanishes at one-loop order, closely paralleling the role played by the Landau gauge in QED$_4$. This observation establishes a direct bridge between LKF transformations, perturbation theory, and multiplicative renormalizability in reduced QED.

We have further analyzed the renormalization properties of the fermion propagator and explicitly demonstrated the multiplicatively renormalizable structure of the wavefunction renormalization in the ultraviolet regime. The resulting expressions exhibit the expected exponentiation of leading logarithms and confirm the consistency of the LKF transformation approach with perturbative renormalization theory.

We then extended our analysis to the nonperturbative regime associated with dynamical chiral symmetry breaking. Employing LKF transformation on numerically generated fermion propagators, we studied the gauge dependence of the wavefunction renormalization and the dynamically generated mass function. Our computation confirms that, although these Green functions themselves are gauge dependent, physical observables extracted from them remain gauge invariant. In particular, the chiral fermion condensate and the fermion pole mass are shown to be gauge independent. 

Overall, our results reinforce the role of LKF transformations as a powerful tool for  preserving gauge covariance in RQED$_{4,3}$ both in its perturbative and
nonperturbative regimes, providing support for the consistency of its symmetry-based framework for planar Dirac materials and related strongly correlated systems.
 
\bibliography{LKF-Reduced-QED}
\bibliographystyle{apsrev4-1}
\end{document}